\documentclass[%
 superscriptaddress,
 amsmath,amssymb,amsfonts,
 aps,
 prl,
 twocolumn, 10pt,
 longbibliography,
 floatfix
]{revtex4-1}
\usepackage[utf8]{inputenc}
\usepackage{graphicx}
\usepackage[sf]{subfigure}
\usepackage{color}
\usepackage{fouriernc}
\usepackage[dvipsnames,svgnames,x11names,hyperref]{xcolor}
\usepackage[low-sup]{subdepth}
\usepackage{MnSymbol}
\usepackage{multirow}
\usepackage{outlines}
\usepackage{etoolbox}
\usepackage{enumitem}

\definecolor{linkcolor}{RGB}{6,69,173}
\usepackage[colorlinks=true,
            linkcolor=linkcolor,
            urlcolor=linkcolor,
            citecolor=linkcolor,
            unicode,
            pdfencoding=auto]{hyperref}

\usepackage[
]{physpack}

\begin{document}

\title{
Local Spectroscopies Reveal Percolative Metal in Disordered Mott Insulators
}
\author{Joseph C. Szabo}
\affiliation{Department of Physics, The Ohio State University, Columbus, Ohio 43210, USA}

\author{Kyungmin Lee}
\affiliation{Department of Physics, The Ohio State University, Columbus, Ohio 43210, USA}

\author{Vidya Madhavan}
\affiliation{Department of Physics, University of Illinois Urbana-Champaign, Urbana, IL 61801}

\author{Nandini Trivedi}
\affiliation{Department of Physics, The Ohio State University, Columbus, Ohio 43210, USA}

\begin{abstract}
We elucidate the mechanism by which a Mott insulator transforms into a non-Fermi liquid metal upon increasing disorder at half filling. By correlating maps of the local density of states, the local magnetization and the local bond conductivity, we find a collapse of the Mott gap toward a V-shape pseudogapped density of states that occurs concomitantly with the decrease of magnetism around the highly disordered sites but an increase of bond conductivity. These metallic regions percolate to form an emergent non-Fermi liquid phase with a conductivity that increases with temperature. Bond conductivity measured via local microwave impedance combined with charge and spin local spectroscopies are ideal tools to corroborate our predictions.
\end{abstract}
\date{\today}

\maketitle

\para{Introduction}
The metal-to-insulator transition (MIT) driven by increasing disorder and the effect of Coulomb interactions on this transition has been a problem of fundamental interest. It is well known that disorder can create a transition from a metallic to an insulating state in both 2D and 3D due to localization effects. In the absence of interactions, all states are localized in one and two dimensions for arbitrarily small potential disorder, while in three-dimensions the MIT occurs at a finite critical disorder strength~\cite{anderson_pr_1958, abrahams_prl_1979}. In the presence of Coulomb interactions, perturbative calculations show an enhancement of localization in all dimensions~\cite{lee_rmp_1985}. However, the idea that disorder can create an insulator-to-metal transition (IMT) is relatively new. The first hint of an IMT in two dimensions due to Coulomb interactions came from the renormalization group (RG) analysis by \textcite{finkelstein_zhetf_1982,finkelstein_zhetfpr_1983,finkelstein_book_1990} which showed that the critical indices for the correlation length and time scales become frequency-dependent and the RG flows take the system to a strong coupling fixed point. This was followed by an RG analysis of a two-parameter theory for long-range Coulomb interactions and disorder in the limit of large number of valleys that found a quantum critical point for the IMT in two dimensions. This theory was successful in explaining experimental data on thermodynamics and transport in high-mobility silicon metal-oxide-semiconductor field-effect transistors (Si-MOSFETS)~\cite{punnoose_s_2005,*punnoose_s_2005_correction}.

\para{}
In the opposite limit of strong on-site repulsion for commensurate filling we have several examples of Mott insulators~\cite{mott_pps_1949} in narrow band systems in which electrons are localized due to strong repulsion with an energy gap to excitations.
The discovery of Mott insulators that can be driven into metallic/superconducting states upon doping has opened the field of competing charge-ordered, spin-ordered, nematic, pseudogap, superconducting and strange metallic phases. In order to understand the emergent behavior, it is important to separate out the effects of adding or removing carriers from simultaneously also increasing disorder. In this regard, gate tuning is a useful knob that tunes only the chemical potential without necessarily adding disorder, as distinct from chemical doping.

\begin{figure*}\centering%
    \includegraphics[width=510pt]{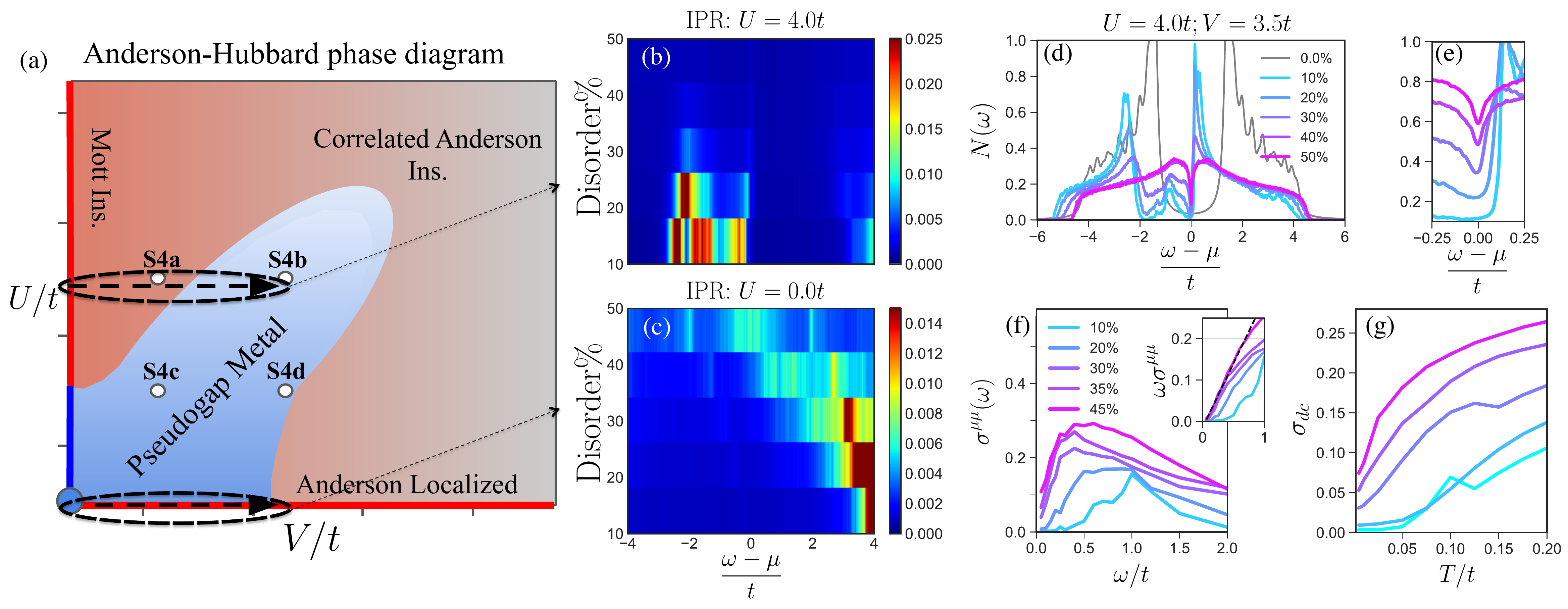}
    \caption{
    (a) Interaction- Disorder $U$--$V$ phase diagram in units of the hopping $t$; $x$-axis captures Anderson localization, $y$-axis is the Mott MIT, and novel metal at intermediate energies. Relative points in phase space where development of conductivity is analyzed to create Fig.\ref{fig:three}(e) (See Fig.S3). Dashed lines represent cuts ($U = 4.0t$ and $U = 0$, $V = 3.5t$ with increasing binary disorder fraction) along which the inverse participation ratio (IPR) is evaluated (b) and (c). IPR $\propto \xi_{loc}^{-2}$ where $\xi$ is the extent of the energy eigenstate.
    (d) Full DOS and (e) states about the Fermi energy offset vertically for clarity (right). Spectral data averaged over 40 disorder realizations for $40\times40$ square lattice with broadening $\delta = 0.0075t$.
    (f) Optical conductivity $\sigma(\omega)$ in units of ($e^2/\hbar$) and $\omega\sigma(\omega)$ (inset) at various disorder fractions. Dotted black line in inset highlights a finite dc conductivity for higher disorder, or correspondingly a linear behavior of $\omega\sigma(\omega)$.
    The optical conductivity exhibits non-Drude behavior with a peak at nonzero frequency.
    (g) dc conductivity as a function of temperature for $30 \times 30$ lattice at $U = 4.0t $ and $ V = 3.5t$.
    The data in (f) and (g) are averaged over 8 disorder realizations.
    }
    \label{fig:diag}%
\end{figure*}

\para{}
Previous experiments have observed a power-law suppression
of the local density of states upon doping the Mott insulator Sr$_3$Ir$_2$0$_7$ with Ru substitution for Ir \cite{Vidya}, which indicates that new states are added at the chemical potential. Ru has been experimentally shown to be an isovalent substitution in Sr$_3$Ir$_2$0$_7$ so the the IMT in this system may be a good example of a disorder induced transition. Transport measurements help determine whether the states are localized or extended, and indeed corroborate a metallic state with a finite resistivity extrapolated to $T=0$ in the Ru substituted compounds. 
Theoretical methods ranging from inhomogeneous mean field theory \cite{heidarian_prl_2004}, DMFT-based approaches \cite{Hofstetter, tmt_dmft_2, tmt_dmft_1, tmt_dmft_0, hof_dmft, dmft_0, cpa_dmft}, to quantum Monte Carlo \cite{QMC_0, QMC_1, qmc_2}, and exact diagonalization studies \cite{ED} of the Anderson-Hubbard model also support the presence of insulator-metal transition (IMT). However the mechanism behind such a transition and the nature of the emergent metallic phase is unclear.

\para{}
In this Letter, we investigate the tension between two localizing tendencies: Mott repulsion and Anderson localization in two dimensions at half filling in the Anderson-Hubbard model, and their roles in driving quantum phase transitions. We investigate the correlations among the local maps of the magnetization, density of states, and the local bond conductivity, for a given realization and strength of disorder. We depict our results in a schematic phase diagram in the interaction-disorder plane in Fig.~\ref{fig:diag}(a) that is based on our inhomogeneous mean field results for the Anderson-Hubbard model. It shows that while both disorder and interactions independently enhance localization of electrons and promote an insulating state, acting together results in a novel metallic phase sandwiched between the Mott insulator at low disorder and a correlated Anderson insulator at high disorder.

\para{}
Specifically, our two key new results are:
\begin{enumerate}[label={(\arabic*)}]
\item Our earlier studies showed that disorder adds spectral weight within the Mott gap resulting in a V-shaped density of states [also shown in Fig. \ref{fig:diag}(d)]. Here we calculate the DC conductivity and show that remarkably the conductivity increases
with increasing fraction of disordered sites and also increases with temperature for a fixed disorder fraction [Fig.~\ref{fig:diag}(g)]. Interestingly, this emergent metallic phase shows a non-Drude response in the optical conductivity [Fig.~\ref{fig:diag}(f)].
\item Local dc conductivity profiles show the formation of conducting bonds in regions surrounding disorder sites, and the emergence of percolating metallic networks.
\end{enumerate}

\para{Model}
The Hamiltonian for the Anderson-Hubbard model is given by
\begin{align}
    H &= -t
    \!\!\!
    \sum_{\langle i, j \rangle, \sigma} 
    \!\!\!
    (c^{\dagger}_{i\sigma}c_{j\sigma} + \text{H.c.}) 
         + U \sum_{i} n_{i\up} n_{i\dn}
         + \sum_{i, \sigma} (V_{i} - \mu) n_{i\sigma},
\end{align}
where $c^{\dagger}_{i\sigma}(c_{i\sigma})$ is the electron creation (annihilation) operator at site $i$ with spin $\sigma$, and $n_{i\sigma} \equiv c^{\dagger}_{i\sigma}c_{i\sigma}$.
$t$ represents the hopping amplitude between nearest neighbor sites, and
$U$ is the onsite electron-electron repulsion.
$V_i$ the onsite disorder potential, treated as binary-alloy disorder:
$p$ fraction of randomly chosen sites have $V_i=V$, and $1-p$ fraction of sites have $V_i=0$.
For each disorder realization, the chemical potential $\mu$ is adjusted to achieve global half filling.
The Hubbard interaction term $U$ is treated at the Hartree-Fock level in terms of the site-dependent spin and charge density fields.
This numerical method has the advantage that it treats the disorder potential exactly, and thus captures the localization physics due to the inhomogeneous potential profile accurately \footnote{For discussion on the numerical procedure, see Supplemental Material.}.

\para{Global properties}
With increasing disorder fraction, the DOS $N(\omega)$ shows an evolution from a Mott gapped insulator to a gapless phase leading to a V-shaped suppression, a pseudogap, at the chemical potential [Fig.~\ref{fig:diag}(d,e)] ~\cite{heidarian_prl_2004,Vidya}. To get some idea of whether these in-gap states are localized or extended, we plot the inverse participation ratio (IPR) $\equiv \sum_{r_i{\alpha}} |\psi_\alpha(r_i)|^4 \propto \xi_{loc}^{-2}$, where $\psi_\alpha$ is the real-space wavefunction associated with eigenenergy $\alpha$ and $\xi_{loc}$ is its associated localization length (zero IPR value corresponds to infinitely delocalized state). The in-gap states at low disorder fraction are bound states that are localized, as shown by the large value of their IPR shown in Fig.~\ref{fig:diag}(b). As disorder regions grow and extend across the whole system, energy eigenstates also become increasingly delocalized, depicting a transition from a Mott insulator to a metallic state.

\para{}
To understand the nature of this emergent phase with a finite density of states at the chemical potential, we evaluate the optical conductivity $\sigma(\omega)$, shown in Fig.~\ref{fig:diag}(f). Starting with a finite gap at low disorder, we observe the gap closing with increasing disorder, consistent with previous theoretical studies \cite{Nirav, dar_thesis}. As the system develops an increasing low-frequency conductivity, the behavior of $\sigma(\omega)$ is non-Drude, with a peak in the conductivity at a nonzero frequency that moves toward lower frequency with increasing disorder.
Rather remarkably, we observe a nonzero dc conductivity that grows with increasing disorder fraction as depicted in low frequency behavior of $\omega\sigma(\omega)$ in the inset in Fig.~\ref{fig:diag}(f).
Beginning at 30\%, the linear behavior of $\sigma(\omega)$ allows us to extrapolate a nonzero dc conductivity $\sigma_{\text{dc}}$ [Fig.~\ref{fig:diag}(g)].
A finite $\sigma_{\text{dc}}$ indicates the onset of a metallic phase in which the conductivity grows with increasing disorder. Similar enhancement of the conductivity was found for disorder chosen from a uniform box potential $[-V, +V]$ \cite{dar_thesis, trivedi_2014, Nirav}, indicating that the emergence of the metallic phase is ubiquitous.

\begin{figure}\centering
    \includegraphics[width=240pt]{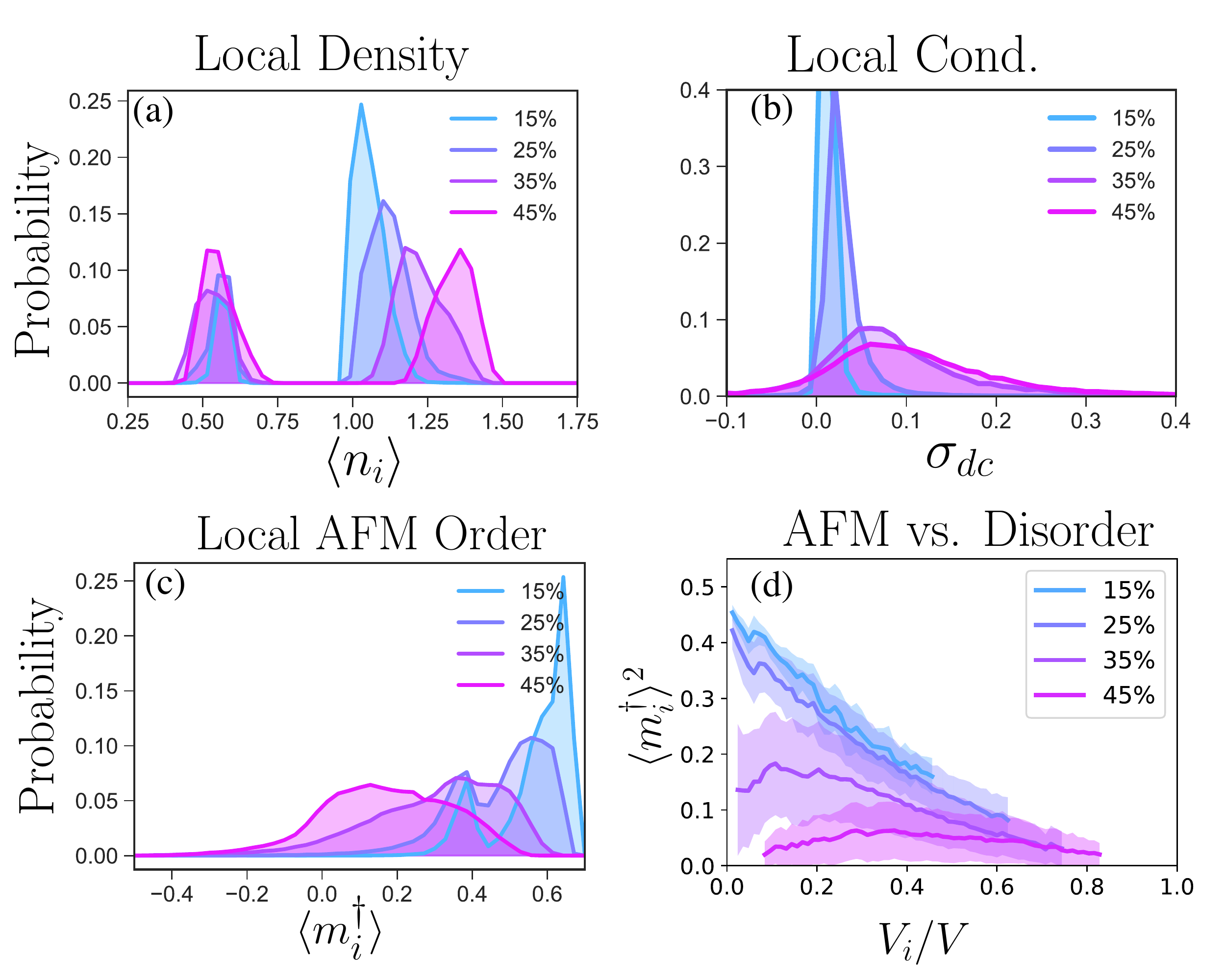}
    \caption{
    (a, b, c) Local charge density, conductivity, and antiferromagnetic (AFM) order parameter ($\langle m^\dagger_i \rangle \equiv (-1)^{x_i+y_i} \langle S_{z,i} \rangle$) as a function of disorder fraction. Data averaged over 40 random disorder realizations at $V = 3.5t$, $U = 4.0t$, $T=0.01t$ for $40\times40$ lattice.
    (b) Local dc conductivity distribution normalized by the total number of sites $N$, extrapolated from the low frequency conductivity.
    (d) Site disorder with Gaussian broadening and its correlation with local staggered magnetization; shading shows the standard deviation.
    }
    \label{fig:two}%
\end{figure}

\begin{figure}
    \centering%
    \includegraphics[width=240pt]{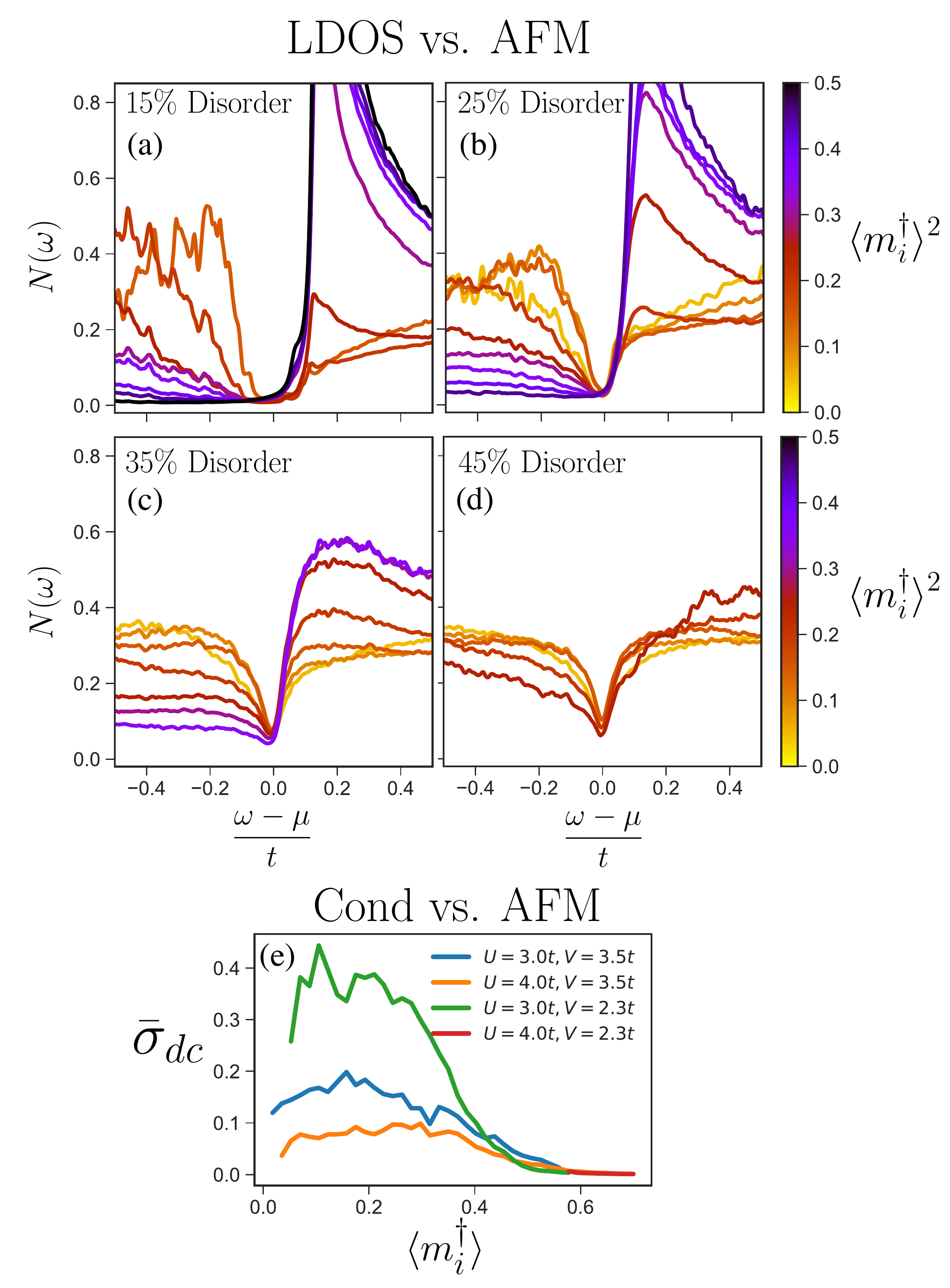}
    \caption{
    (a-d) Inhomogeneity in LDOS and dependence on $\langle m_i^\dagger \rangle^2$. Data collected from 40 random disorder realizations on a $40 \times 40$ lattice, $(U,V) = (4.0t, 3.5t)$, 
    (e) Average local bond conductivity as a function of $\langle m_i^\dagger \rangle$ taken from 8 disorder realizations for 15\%, 25\%, 35\%, 45\% on lattices of sizes $26\times26$.
    See Fig.~S3 for full conductivity distribution vs ordering.
    }
    \label{fig:three}
\end{figure}

\medskip

\para{Local properties}
Insight into how metallicity arises as a result of the competition between disorder and interactions is captured by the distribution of local quantities: antiferromagnetic (AFM) order parameter $\langle m^\dagger_i \rangle \equiv (-1)^{x_i + y_i} \langle S_i^z \rangle$, LDOS, and transport characteristics $\sigma^{\mu\mu}_{ij}$, as we discuss below.

\para{(a) Local magnetization}
For positive potential $V$, it becomes energetically unfavorable to occupy the disorder sites, leading to a reduction in the local moment and charge density on disorder sites. The bimodal charge density distribution shown in Fig.~\ref{fig:two}(a) depicts disorder sites with relatively fixed mean occupation while non-disordered sites slowly transition away from unit filling, initially only impacting nearest neighbor sites.
The spin ordering distribution echoes this nearest neighbor to disorder behavior. In Fig.~\ref{fig:two}(c), the distribution of AFM order is sharply peaked close to the maximum value at low disorder fractions, and becomes broader and shifts toward zero as the fraction increases, indicating a transition from a uniform AFM phase toward a nonuniform paramagnetic phase.
Fig.~\ref{fig:two}(c) shows a reduction of the local moment on neighboring sites as the occupation increases beyond unit filling: Sites are screened from the effects of disorder at low disorder fraction, seen by the disparate peak and slightly perturbed sharp AFM profile for $15\%$.
As the density of disorder in local regions grows, more charge occupies neighboring sites and neutralizes spin ordering. 
We correlate the magnetic ordering with local disorder density by smoothing the original disorder potential to create an effective disorder profile [Fig.~\ref{fig:two}(d)]. At low fraction $\langle m_i^\dagger\rangle^2$ decreases linearly with the degree of disorder, so initially $V_i$ only has localized effect on interacting sites. As the disorder throughout the full lattice grows, the impact becomes increasingly non-local where sites away from disorder become paramagnetic.

\begin{figure*}
    \centering%
    \includegraphics[width=250pt]{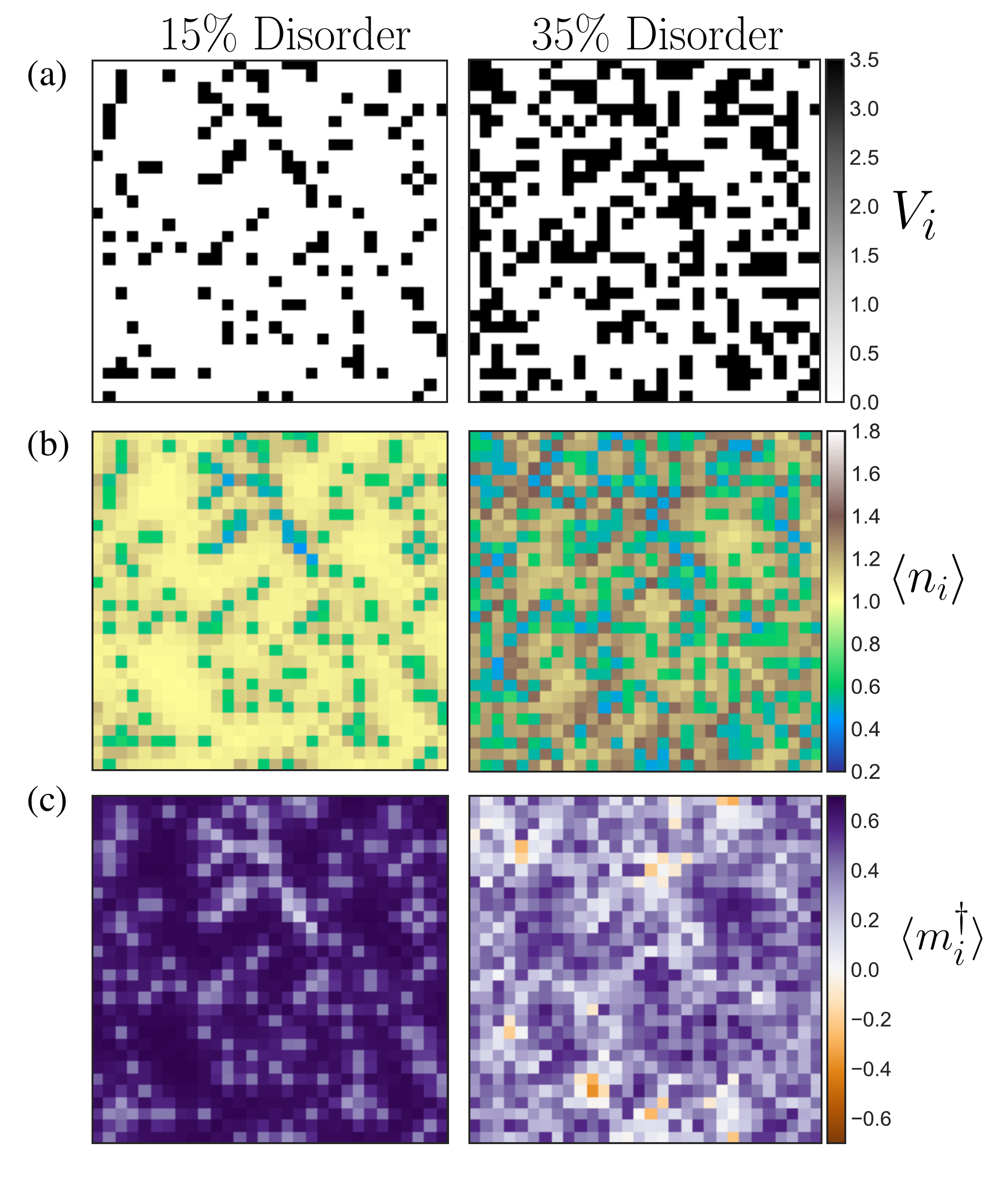}
    \includegraphics[width=230pt]{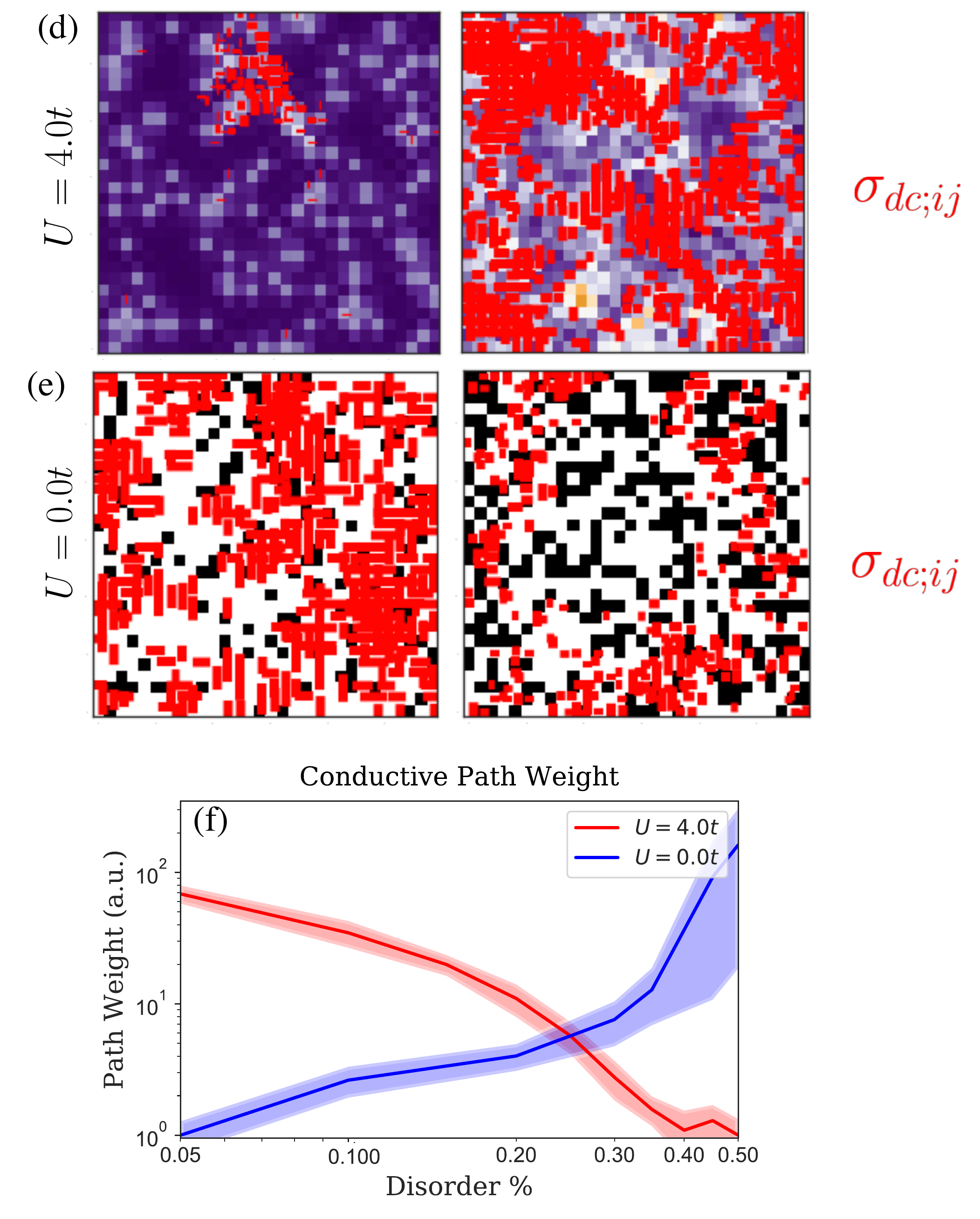}
    \caption{
    (a,b,c) Real space profile for two representative disorder realizations  ($15\%$, $35\%$) and corresponding charge density and AF magnetization for a $30\times30$ square lattice and $U=4.0t, V=3.5t$. (d) Local staggered magnetization map superimposed with local bond conductivities (red lines), obtained at low frequency and low temperature ($\omega = 0.01t, T = 0.005t$) for two disorder realizations.
    Lines in $35\%$ (right) are roughly 10--20 times more conducting to similar intensities in 15\% (left).
    (e) Recreation of figures (d) with interactions turned off and identical disorder profile.
    (f) Shortest distance path for conductivity network constructed from the conductivity profile. Edges between sites $(i,j)$ assigned weight proportional to $1/\sigma_{ij}$ and minimum path algorithm used to analyze network. Path weight/distance is in arbitrary units but proportional to $N^2/\sigma_{ij}$, normalized to minimum path weight (50\% for interacting, 5\% for noninteracting cases). Constructed from $L$ paths for each $L \times L$ random disorder realization ($L = 30$, and 10 disorder profiles).
    }
    \label{fig:four}
\end{figure*}

\para{(b) Local bond conductivity}
The distribution of local conductivities (current-current correlator between bond $(i,j)$ and all $(k,l)$ bonds) in Fig.~\ref{fig:two}(b) shows that mostly clean systems have a local dc conductivity distribution that is sharply peaked near zero, as expected for a Mott insulator. As disorder increases, the bond conductivity distribution broadens and the mean increases. 
Thus bonds become increasingly conducting and sites increasingly paramagnetic as the system becomes more disordered.

\para{(c) Inhomogeneous LDOS}
The magnetization and conductivity hint at an inhomogeneous nature of the emergent disorder-driven metal. We ask how does this local non-uniformity promote charge transport in a Mott insulator. The first insight into how a metal emerges with increasing disorder comes from local spectroscopic analysis. Fig.~\ref{fig:three} depicts the local density of states averaged over sites with different ranges of magnetic order. For low disorder fraction [Fig.~\ref{fig:three}(a)], regions with high AFM order exhibit a Mott-gap around the Fermi energy with almost no states below $E_F$. For moderate disorder, [Fig.~\ref{fig:three}(b,c)] show that Mott physics is preserved in magnetically ordered regions, while increasingly disordered regions have enhanced spectral weight within the Hubbard gap with the formation of a V-shaped pseudogap. Such spectroscopic dependence on disorder has been observed experimentally in Mott insulating materials, where the pseudogap behavior is enhanced near impurity atoms \cite{Vidya}.

\para{(d) Correlation between local moment and conductivity}
Extending the previous discussion on the correlation of eigenstate delocalization and closing spectral gap with reduced magnetic ordering, to low frequency conductivity in Fig.~\ref{fig:three}(e), we show that transport and magnetic order are anti-correlated: The less magnetically ordered the region, the more conducting. Introducing few disorder sites decreases the magnetization and drives charge mobility on these sites, see Fig.~\ref{fig:three}(e).
Maximal conductivity occurs at nonzero magnetic ordering in each curve, suggesting that weak correlations between sites is crucial for promoting mobility.

\para{} 
In Fig.~\ref{fig:four} we present a real space picture of two representative disorder realizations at $15\%$ and $35\%$ to show how disorder breaks down an initial Mott insulating system. Fig.~\ref{fig:four}(a-c) provide a spatial map of the random disorder potential and its effect on charge and spin.

\para{}
To relate disorder and interactions with their effects on the local magnetization and bond conductivity in real-space, we overlay the magnetization profile for a representative disorder realization with the most conducting bonds (seen in red) in Fig.~\ref{fig:four}(d).

\para{}
The contrast with the non-interacting system (Fig.~\ref{fig:four}(e)) is remarkable: While disorder reduces conduction as expected in the non-interacting case, the behavior is quite the opposite for the interacting system, where one observes pockets of enhanced conductivity localized around a small fraction of disorder  sites [$p=15\%$, Fig.~\ref{fig:four}(d, left)], expand to a percolating cluster at larger disorder fraction [$p=35\%$, Fig.~\ref{fig:four}(d, right)].

\para{(e) Formation of percolating metal}
To exhibit the percolative nature of transport, we construct a network with each site as a node and $\rho_{ij}=1/\sigma_{ij}$ as the bond weights. The minimum series resistance $\rho_{min}$ to connect the two ends of the system by the shortest path is obtained by a weighted path analysis of the conductivity graph[Fig.~\ref{fig:four}(f)].
The addition of 10\% disorder leads to a factor of 2 decrease $\rho_{min}$ while adding 30\% disorder leads to a decrease by nearly two orders of magnitude.
Above 30--35\% disorder the cost remains constant as expected above the percolation threshold. The percolative nature of disorder we have characterized adds to current theoretical descriptions \cite{perc_doping, perc_doping_hf, local_nmr_0, local_nmr_1} and experimental characterization of disorder in materials \cite{dis_perc_exp}.

\para{}
In conclusion, our results capture the previously unexplored local properties of the intermediate pseudogapped metallic phase in the half-filled Anderson-Hubbard model. This work illuminates the interplay between disorder and interactions in the simplest fermionic model, and provides an avenue for understanding how non-Fermi liquid behavior arises at the microscopic level.

\para{}
The conductivity maps provide theoretical predictions for increasingly powerful spatially resolved spectroscopic techniques such as microwave impedance microscopy, 4-probe STM, and LC-AFM, which only recently been used to study local conductivity profiles on 100 nm down to atomic resolution \cite{mimit, mim_0, lc-afm}. Recent breakthroughs in cold atom experiments now allow for resistivity and optical conductivity experiments, where the highly tunable nature of these experiments provide an ideal testing environment for the results we present here \cite{cold_opticond}.

\begin{acknowledgments}
We acknowledge useful discussions with J. O'Neal and N. D. Patel.
J. S. and K. L. were supported by the National Science Foundation Grant No.~DMR-1629382.
\end{acknowledgments}

\bibliography{references}

\end{document}